# Finite element modelling and investigation of the interaction between an ultrasonic wave and a discontinuousinterface

Vipul Vijigiri*[a] [b], Cedric Courbon[b], Guillaume Kermouche[c], Juliette Cayer-Barrioz[a]

[a]Laboratoire de Tribologie et Dynamique des Systmes, Ecole centrale de Lyon, 36 Avenue Guy de Collongue, Lyon, France

[b] Laboratoire George Friedel, Ecole nationale suprieure des mines de Saint-tienne, 158 Cours Fauriel, France

[c]Laboratoire de Tribologie et Dynamique des Systmes, Ecole nationale d'ingenieurs de Saint-Etienne, 58 Rue Jean Parot, France

**Abstract**

When two surfaces are brought into contact and slide against each other, junctions are formed at the interface. The dynamics of formation, rupture and evolution of these junctions governs the tribological response of the macro-contact. Getting insight on the real behavior of these junctions is a challenging task. Theory states that contacts and asperities are continuously altered in two bodies due to applied pressure, which increases the number of active contacts. To addresses such altering interface conditions, wave propagation through tribological interface by means of the development of a numerical model is proposed. The proposed method is used to study and relate crucial parameters like stiffness, contact width, number of asperities that form a basis for an interface.

## 1 Introduction

Contact theory had been of great importance past few years to study the interface changes with respect to the crucial tribological aspects i.e., temperature changes, wear and material transfers etc. [1]. Non-destructive testing of such tribological contacts has always been growing research with use of innovative methods like ultrasound. In-situ accessing cracks, defects using ultrasound wave propagation through the contacting bodies is an efficient method characterized by reflection and transmission wave through a crack or a tribological interface. Although, such measurements are sensitive to frequency, amplitude of the wave, type of contact and its properties. In general, Tribological contacts are formed when two bodies are pressured on to one another creating an imperfect contact interface which is never perfectly coupled. Microscopically, they have peaks and valleys due to the surface roughness as seen in figure 1. The points at which the surfaces in the contact intersect, they contribute to the real contact area. This real contact area is pretty small compared to the apparent contact area in magnitude. Real contact area is calculated from the interface based on the number of contact points and diameter of all the asperities in contact. Although, the real contact area is never consistent. To access this real contact area, several analytical models were proposed that built the relationship between the wave propagation parameters to contact area, but it is still a challenging procedure to work on rough contacts.

---

*Email addresses:* `vipul.vijigiri@enise.fr` (Vipul Vijigiri*[a] [b]), `cedric.courbon@enise.fr` (Cedric Courbon[b]), `kermouche@emse.fr` (Guillaume Kermouche[c]), `juliette.cayer-barrioz@ec-lyon.fr` (Juliette Cayer-Barrioz[a]).

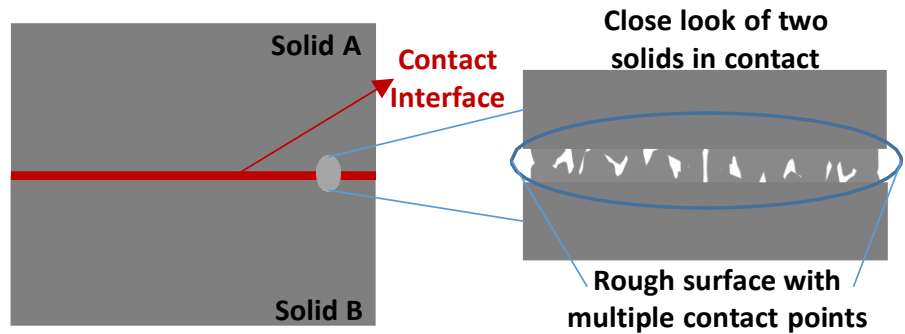

Fig. 1. Tribological contact representation

The real contact area is very difficult to access or measure due to being small. Few methods like a) thermal b) ultrasonic, c) optical, d) resistive etc. are previously researched in reference to the contact theory [2], [3], [4], [5], [6], [7] but fall short due to their disadvantages. Ultrasound on the other hand is popular method like optical but it could also be applied and as well could propagate through any material. However, the existing literature on ultrasound in tribology focus more on the theoretical validation and experimental analysis of wave propagation which have their limitations in-terms of either geometry or complexity in analyzing micro scale contacts. There is also lack of literature on finite element modelling of wave propagating through such contacts and reviewing the contact theory. Use of ultrasound transmission in the evaluation of dry contacts initially was developed by [Kracter,1958] who investigated the performance of wave transmission through an interface with varying load which was later validated in detail by Greenwood (1966) [8], Kendall and Tabor(1971) [9], Yoshihisa MINAKUCHI [10] etc. Work done by Kendall and Tabor gives an interesting insight about ultrasonic use in accessing contacts. They conducted various studies on wave propagation through a soft material when pressed between two solids, through a continuous and discontinuous interface, through single and multiple contacts and through stationary and sliding contacts. They conducted series of experiments initially on cylindrical disks pressed onto one another and in between are the asperities of soft rubber (0.1 mm in size each). These disks are then distanced (0 to 3 diameters) apart for each experimentally to study the change in stiffness (calculated based on the radius of the embracing contact). Finally, the relation of stiffness was built to the number of asperities on two conditions as seen in figure 2.

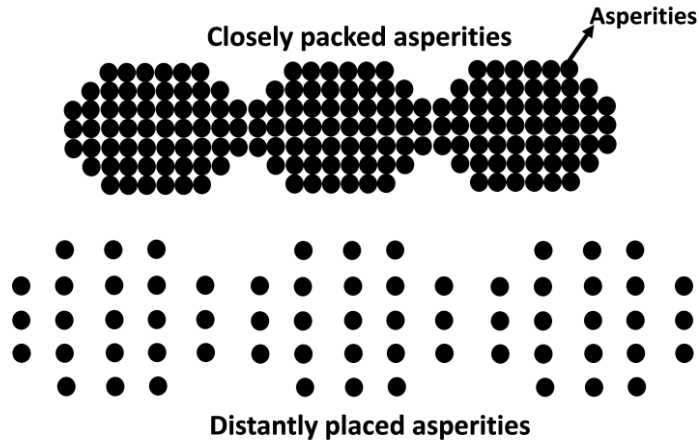

Fig. 2. Asperity distribution types in a contact when they are far apart and when they are closely placed

Later, transmission coefficient was related to change in diameter of a constriction and a disk, a disk was placed, and its width is increased (1-10 mm scale) between two compressing steel bodies for every simulation and resulting transmission coefficient was compared. In both the cases, the analysis is very promising with a few limitations that, in the first case, the asperity is limited to rubber and in the scale of mm, In the second case, transmission coefficient was compared to one single soft disk with a growing diameter and concluded up on that, transmission coefficient is proportional to diameter of the disk and then to the stiffness (Stiffness is then proportional to the diameter of the asperity). The main question would then be, if is the hypothesis still applicable to asperities with different materials and sizes (in microns or below), also a standard explanation or result was not given or concluded relating transmission coefficient to stiffness. Polijaniuk, A. and Kaczmarek, J. (1993) [11] assumed that real contact area is equal to the square times the reflection coefficient. Baik, 1984 [12] dealt majority on ultrasonic scattering from imperfect Interfaces using A Quasi-Static model, His work covers interfaces consisting of strip, circular, elliptical cracks and random distribution of contacts. He further commented on Kendall and Tabor's work that there is a lack of observations concerning measuring real contact area based on stiffness and number of asperities proving the point of further study. N.F.Haines(1980) [13] worked on wave transmission and reflection through contacting surfaces and also on the wave scattering from the contacting surfaces and cracks. Drinkwater (1994) [14], Nagy (1992) [15], S. Biwa (2005) [16] dealt majority on the wave propagation parameters and stiffness considering theoretical assumptions based on spring theory. Other researchers, Rober Dwyer Joyce (2005) [17], Bushan (1998) [18], Krolikowskiand Szczepek 1991 [19], worked exclusively on the experimental or analytical investigation of the interface or the real contact area, developed theory on stiffness Based on the spring model using quasi static approach. It is to be noted that there is an interesting disagreement in the literature on how the wave propagation parameters are related to either stiffness or real contact area. Models or theories that are developed previously have limitations interns of the either the type of interface they considered with fewer contact points and conditions. Also, theory of stiffness and true area of contact stands true in certain conditions but as we increase the no of contact points, effect of the wave propagation in the asperities changes drastically. This could attribute due to change in contact interface in factor of microns.

The goal of this paper is to FE model and study the relationship between contact interface to try to answer few aspects that are not covered previously by the researchers [9]. Lot of work has been done on the finite element modelling of either the surface topology or the asperities in contacts [20] [21] but not using ultrasonic wave propagation through contacts. In this paper, finite element modelling (in micro scale) on the use of ultrasound pulsed on to a solid-solid interface is proposed. Contact interface is altered accordingly to evaluate the parameters that effect the wave propagating through. A detailed study is conducted on the interface configuration and its dependency on crucial parameters (stiffness, contact ratio, contact width, number of asperities) relating stiffness to the ultrasonic transmission coefficient and to interface configuration.

## 2 Theoretical Literature

Generally, in tribological contacts tested using ultrasound, the quality of the contact is accessed through the reflection coefficient and transmission coefficient respectively, which is relatively dependent on real contact area. Real contact area is very small compared to the apparent contact area. It is due to existing surface roughness and small asperities that form due to friction and sliding. Assessing such contacts is not always easy as the smaller they get, visible area per contact reduces. Theoretical assumptions of asperity as a spring were made and finite element models were developed concerning the on-line measurement of such contacts. FE modeling and mathematical formulation of the propagation of ultrasonic waves was first formulated by [22], [17] and later by [23]. The authors considered an incomplete interface and characterized it as a spring layer as seen in figure 3a. As stated by Baik, 1984 [12] before, real contact could be considered as a spring as seen in figure 3b.

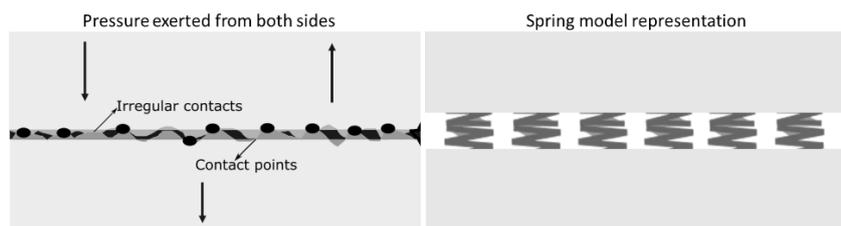

Fig. 3. (a) Tribological contact representation (b) Spring model representation

The quasi-static model that theoretically relates the reflection coefficient to impedance of the two bodies in contacts is given as

$$R_c = \frac{(Z_1 - Z_2 + i\omega(Z_1 Z_2 / K))}{(Z_1 + Z_2 + i\omega(Z_1 Z_2 / K))} \tag{1}$$

Rc is the reflection coefficient, z1, z2 are the impedances of two interfacing bodies, $\omega$ the wave angular frequency and k is the stiffness. The equation can take a simpler form if the contacting solids are made of the same materials, i.e., same acoustic impedance (z1=z2=z), equation 4.

$$|R_c| = \frac{1}{\sqrt{1 + (\frac{2k}{\omega z})^2}} \tag{4}$$

The relationship between reflection coefficient and stiffness forms the basis for much interesting analysis as an interface can be accessed for different loading conditions based on the amount of reflection recorded for each amount of load applied. The reflection and transmission of waves at the interface of two bodies is dependent not only on their characteristic acoustic impedance, stiffness of the interface but also on the frequency of the pulse. In a real contact, the wave propagation occurs in micro scale. Each single contact in the model exhibits the reflection and transmission phenomena. All these interact with one another resulting in change of the characteristics of wave propagation. Their model is quite simple with series of circularly identical contacts are placed in between two compressing bodies. The stiffness,K for such interface can be given as,

$$K = N^b . K_0 \tag{5}$$

$k_0$ is the stiffness for each single asperity and b is a numerical variable that varies between 1 for when the N contacts are spaced so far to have negligible effect and 1/2 when N contacts are closely or tightly placed such that one has considerable effect on other. Perfect match or a perfectly closed interface has impedance close to zero or high stiffness allowing complete transmission and as if the air gaps are increased in the interface, transmission coefficient will reduce to zero. In both [9] and [22] research works, stiffness of the whole interface separating two solid bars was proved to be dominant parameter that controls the wave reflection and transmission.

## 3   Finite element Model geometry and Interfacial configuration

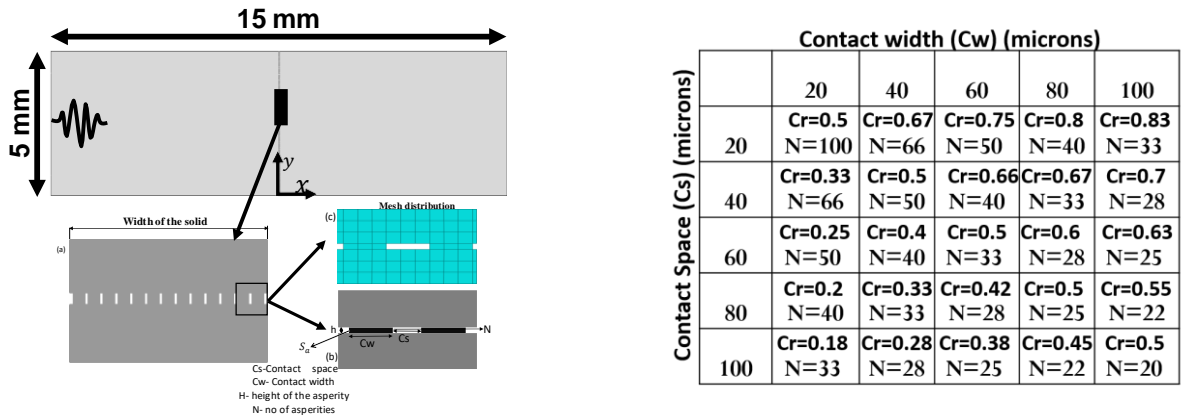

Fig. 4. (a) Tribological contact representation (b)Table of contact width, contact space and asperity combination of the interface

FE modelling is realized using ABAQUS/Explicit. Element type used is a first order 4 node bilinear solid element (CPE4R). In the model, aluminum has a 15 mm length/Height and 5 mm width geometry seen in the figure 4. Young's modulus, Poisson's ratio, density, impedance is shown in the table below.

| Density $\rho (kg/m^3)$ | Young's modulus E(GPa) | Poisson's ratio |
|---|---|---|
| 2700 | 69 | 0.33 |

Table 1. Properties of the material (Aluminum) used in the FE model

The simulations were conducted with different contact or interface configurations characterized by the contact width (Cw) (20-100 microns), Contact space (Cs) (20-100 microns), and number of asperities (N). For number of asperities (N) from 20-100, corresponding contact ratio (Cr) is indicated as seen in figure 4. The geometry consists of quadrilateral mesh on a displacement element type. The mesh value is fixed to 20 microns in all simulations to maintain the mesh stability and to make sure the simulations are consistent. The solver employed for the simulations is explicit time dependent, dynamic temperature-displacement solver. The evolution of displacement along time of the global pressures is analyzed, relating it to the local interaction phenomena at the interface and global behavior of the solid-solid interface. Ultrasonic tone burst with 10 MHz frequency, 6 cycles and 1E+7 amplitude is applied as a load boundary to the element and then the output is measured on both side of the element. The tone burst is formulated as,

$$A = A_0(1 - \cos(2\pi f \frac{t}{N})) * (\sin(2\pi f t)) \tag{6}$$

$A_0$ here is the amplitude. Input is a hanning windowed tone burst applied along the blue line in figure 5a. The excitation waveform and its first harmonic reflection and transmission are observed in the figure 5b. The time signal is Fourier transformed with in a time frame to extract the amplitude of Excitation, reflected and transmitted signals as seen in figure 5c.

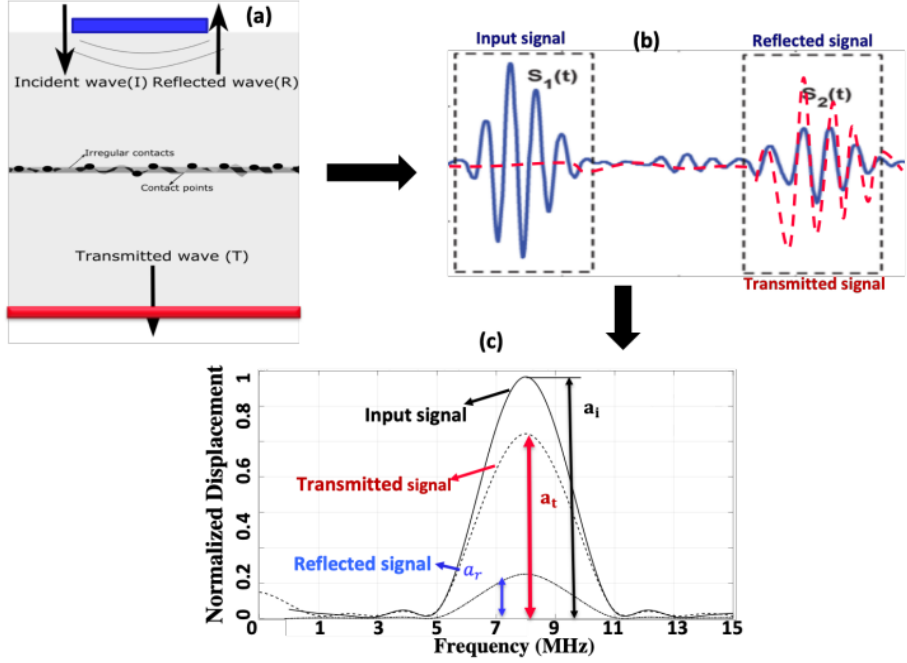

Fig. 5. (a) Solid-Solid interface (b) Time domain representation of input, reflected and transmitted waves (c) FFT or frequency domain representation of input, reflected and transmitted waves.

Reflection and transmission coefficients as seen in equation 7 are calculated from ratio of the maximum displacement amplitudes extracted from the figure 5c. In the above figure S1(t) is the time signal representing input pulse and S2(t) is the time signal representation of the reflecting and transmitting parts. Also, ai, at, ar are amplitudes of the input, transmitted, reflected signals that are recoded. In general, the sum of both the coefficients equate to unity. The transmission coefficient is extracted from the 80 nodes on the right boundary of the solid and summed, resulting an estimate. Reflection coefficient is extracted from 45 nodes on the left side of the boundary.

$$R_c = \frac{a_r}{a_i} \tag{7}$$

$$T_c = \frac{a_t}{a_i} \tag{8}$$

## 4 Mesh sensitivity study and consistency test on a single solid body without interface

The level of accuracy of a wave propagation model is greatly influenced by the mesh density and time step. This section will project upon the influence of 2 mesh distributions with the aim of quantitatively defining a suitable mesh distribution that will be further used in later simulations. The mesh distribution is scaled from a coarse 200 to a fine 20 microns mesh in this sensitivity analysis. For the complex micro scale structures the mesh density must be very high as in time domain wave propagation, for each node, displacements are calculated and averaged. Although time consumed to process is very high, wave propagation within the solid is depicted in precise detail. The quadratic element mesh distribution was used considering its advantage of meshing small scale structures (Abaqus manual, 2014). The effect of the mesh distribution/quality was first investigated by applying a tone burst on one side of the solid, whereas displacement amplitudes were extracted on both the sides. Following figure shows the wave propagation modes for two mesh types.

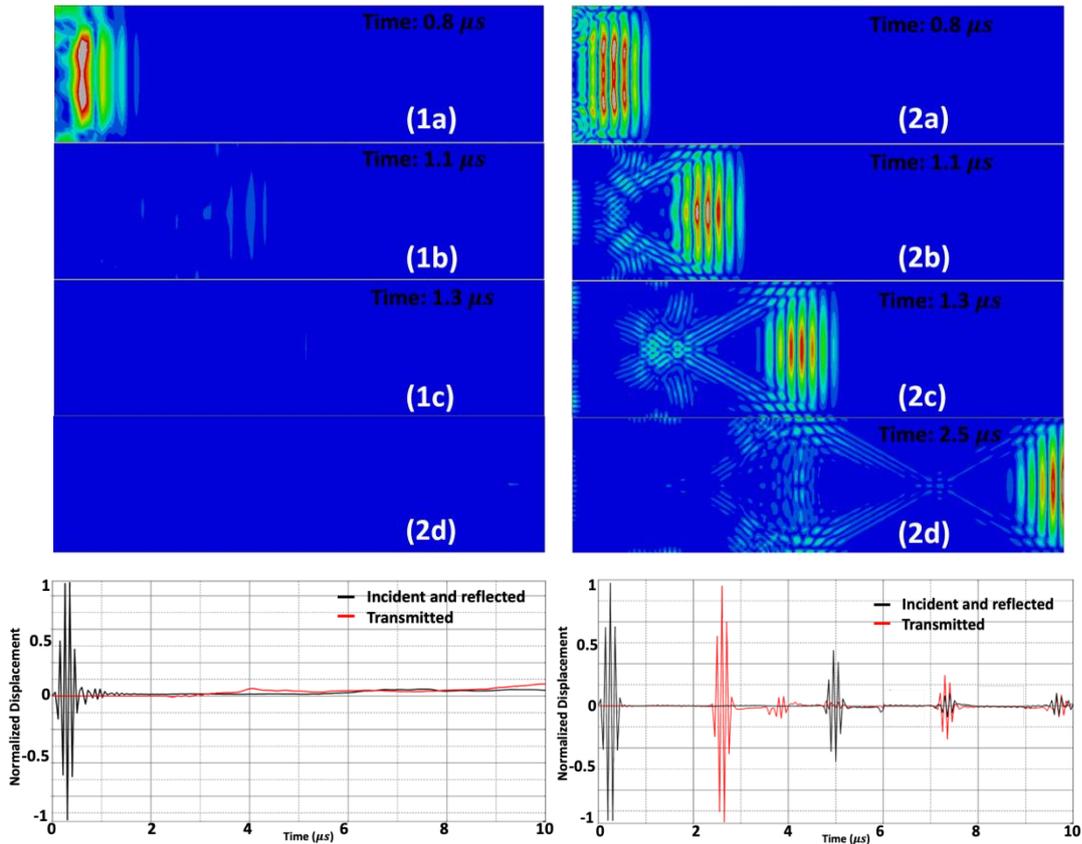

Fig. 6. Time lapse of wave travelling as it reaches the right end of the solid for (1) coarse mesh distribution (2) Fine mesh distribution Displacement magnitude (a) plotted for the coarse mesh distribution (b) plotted for the fine mesh

From the results shown in figure 6, the propagated wave is seriously affected by the mesh distribution as it reaches the right end of the solid, whereas the scale for the 2 simulations is similar. There are two obvious observations, one is that the wave itself, as it starts propagating, has different displacement patterns. On the other hand, as it propagates through, the wave energy when using the coarse mesh is completely lost with almost zero transmission along the field. When using the fine 20-micron mesh, in the figure 6(2d), the peak amplitude is almost maintained until the end of the solid despite the slight beam spreading and scattering on the side walls. On the contrary, the figure 6 (1c) proves that the wave pattern is rapidly lost leading to a strong decrease in the wave amplitude. In practice, this will not be the case as based on the length of the solid, there could be a clear wave travelling along and reflecting unless altered due to attenuation or scattering.

Figure 6 (a) presents the total displacement recorded in the propagation direction on both sides of the solid. In figure 6 (b), it can be noticed that the wave is almost fully transmitted to the right end of the solid whereas the first echo recorded on the left at 5 s is weaker in amplitude. When using a coarse mesh distribution, the whole signal intensity is lost over the few propagation steps which is noticeable. Based on the meshing rule, these simulations prove that smallest mesh size are necessary to reach an accurate performance of model and vice versa. This concludes that change in mesh density does affect the wave propagation efficiency and also introduce numerical inaccuracies. From multiple analyses on different mesh distributions, an efficiency curve versus mesh size was drawn as seen in figure 7.

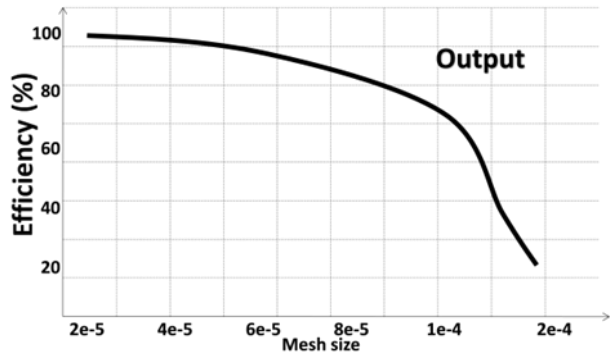

Fig. 7. Mesh distribution size and its efficiency curve

Efficiency here is defined as the ratio of the incident amplitude to received amplitude given as seen in equation

$$\eta = \frac{Output\ amplitude}{Input\ amplitude} * 100 \qquad (9)$$

Based on the resulting curve, it clearly shows that when increasing the mesh size, the efficiency of the output reduces to nearly 10 percent. This concludes that 20 microns is a very good mesh size than can be adapted to the present model. Usually, the mesh size has to be 1/10 elements per wave length to simulate the wave propagation and in this case, the mesh size is 1/25 elements per wavelength.

## 5 Wave propagation mode shapes at different time intervals

The wave propagation through the contacting bodies is always interesting to show and it gives a clear overview of the wave propagating over time. Figure 23 show all the color plots of the wave when it is first initiated, propagated through the interface and as it reached the end. Within the interface, the part of the wave can be seen reflecting and part being transmitted through. The time plot is also seen in figure 24 which shows the first initial pulse that is fed to the solid and the second transmitted and reflected pulses. Harmonics are periodically seen until the amplitude is reduced near to zero (The plot corresponds to one contact / width-60 microns and number of asperities-33).

## 6 Random contact distribution

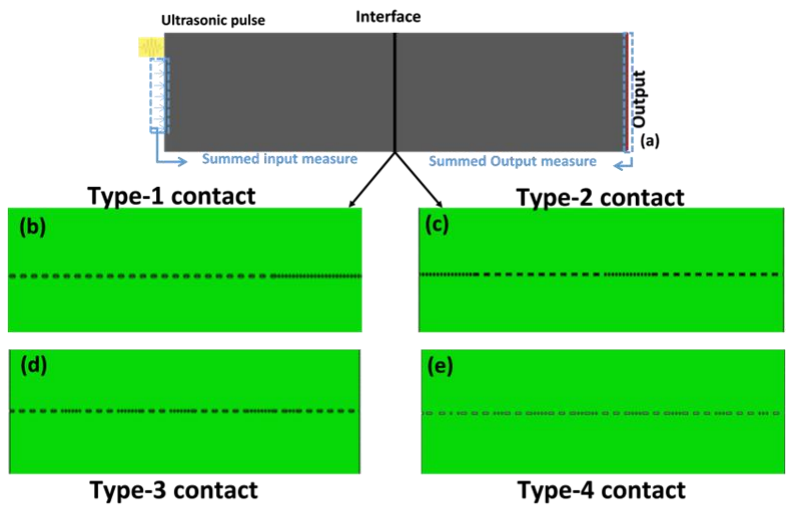

Fig. 8. a) Solid-solid contact interface with the interface shown in black strip (b) (c) (d) (e) four different contact distribution types

In the previous section, a rectangular shaped asperity with a varying size was uniformly distributed to assess the effect of a change in the number of contact spot only. The question now could be, how does the wave behave for a given number of contact asperity but with a variable spatial distribution? In this modelling part, the size of solids remains same as specified in previous section, but the interface configuration is designed with a constant N=50 number of asperities with spacing of 60 microns and 20 microns in combination as seen in figure 8. All the cases result in the same contact ratio, Cr of 0.5

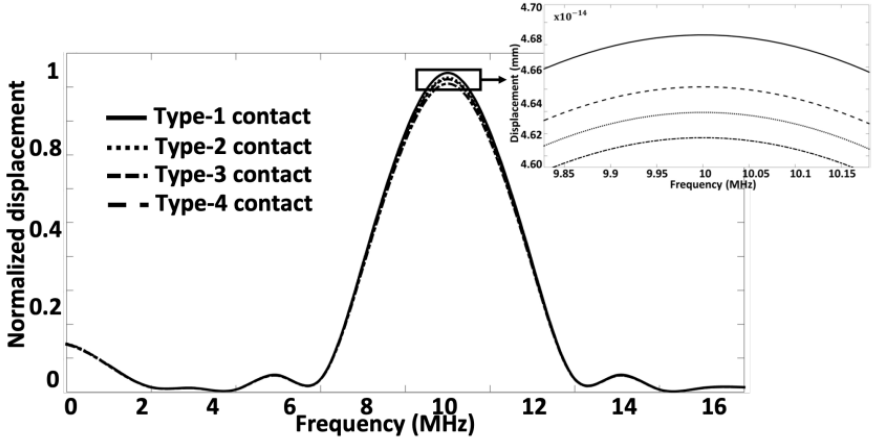

Fig. 9. FFT plot (Transmission) of the four-time domain plots corresponding to irregular interfaces

Figure 9 shows the FFT plots for four type of irregular interface distributions types. It can be observed that in all the cases, the amplitude remains almost unaltered with slight difference in levels. This tends to show that the interface at such a scale will not have a drastic effect on what is measured at the transmission end. This concludes that change in interface configuration for a given contact ratio will have minimal impact on the simulation capacity.

## 7 Effect on the wave transmission with change in width of the asperity

Considering a configuration with a 10-micron thick asperity, the study is focused on assessing the evolution of the reflection coefficient to change in width of a single asperity. Geometry in this case is similar to the previously defined model as seen in 4, except that the width of single asperity is changed from 0.1 to 3 mm. Figure 10 shows the corresponding images of wave as it partially reflects and transmits at the asperity.

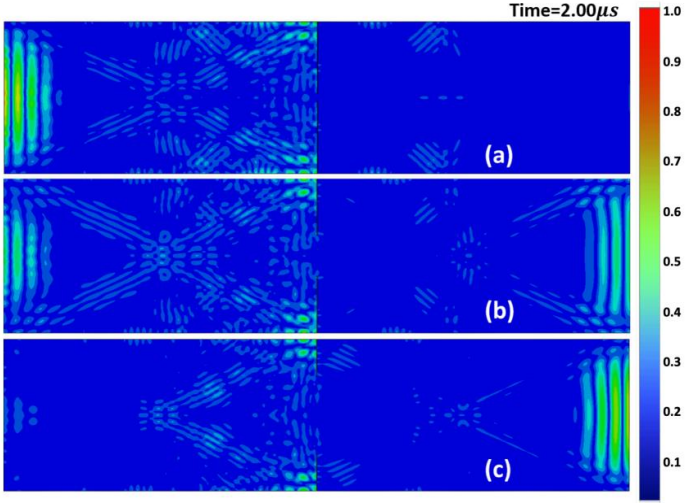

Fig. 10. Wave propagating through different asperity width (a) 0.1 mm where all the wave is reflected back (b) 1.5 mm where part of the wave is transmitted and part reflected (c) 3 mm where most of the wave is transmitted (results are extracted at same time frame)

It is noticed that as the width of the asperity increases, amount of wave that is reflected reduces to near zero (once the asperity reaches 5 mm) and transmission increases gradually. In such cases with increased width, the fronts cannot interact with free spaces between the two contacting bodies which results in increased propagating wave. Figure 11 shows the relationship between the width of the asperity and the transmission coefficient being proportional to the width of the asperity.

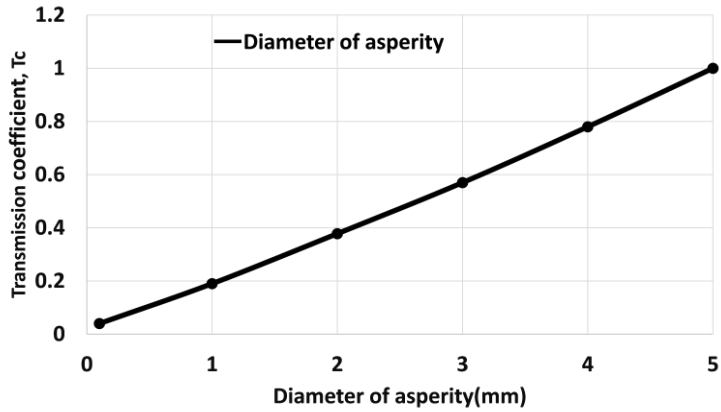

Fig. 11. Relationship between width of the asperity to transmission coefficient

This appears to be in agreement with the work presented by [9] confirming the consistency of the modelling approach and the potential use for further analyses.

## 8   Relationship with the stiffness of the interface

It was stated in [9], that transmission is in fact proportional to the stiffness of the interface resulting from the increase of the asperity width. In order to verify this relationship, numerical simulations were carried out to assess the stiffness of each asperity size configurations. A simple 2D Axi-symmetric single asperity punch model is developed and studied in terms of changing contact width. The model is simulated in static, general and the element type is a 4-node bilinear plane stress. A short displacement of -0.1 is applied on the top boundary x and a symmetry boundary condition is applied in both x and y directions. Reaction Force is measured as output over time. A sketch of the geometry is shown in figure 12 with the corresponding input data in table 2.

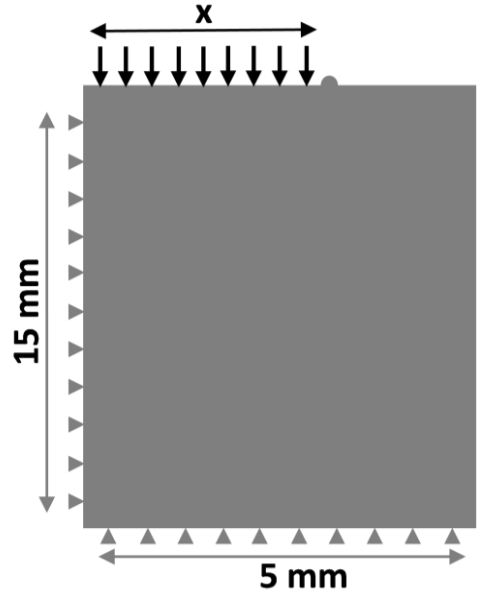

Fig. 12. Schematic representation of the model assumed with load and the boundary conditions defined

| Part definitions | Part type | 2D deformable |
|---|---|---|
| | Material | Aluminum |
| Load | Step | Static general |
| | Displacement, Y | -0.1 |
| | Output | Force |
| Boundary conditions | Symmetry | X direction |
| | Symmetry | Y direction |
| | Signal window | Hanning |
| Meshing dimensions | Element size (mm) | 0.25 |
| | Element type | Explicit/Plane stress |
| | Mesh control | Quad dominated/Free |

Table 2. Table corresponding to the load, boundary conditions, mashing rules implemented in this section

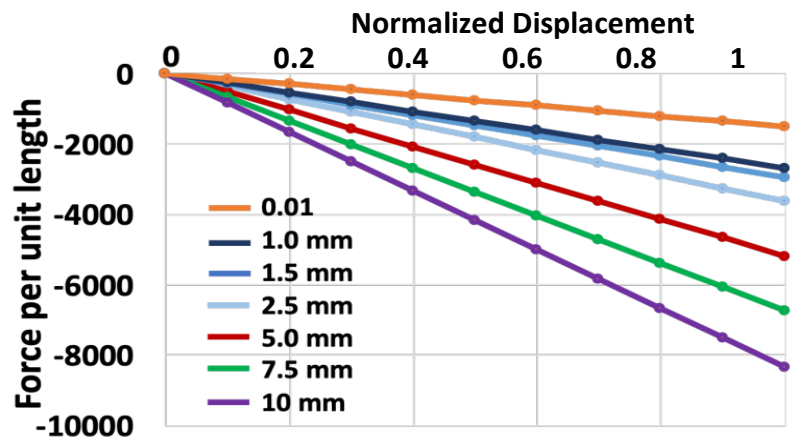

Fig. 14. Reaction force versus displacement for different asperity width configurations

Figure 14 shows the reaction force plotted versus the punch displacement. The stiffness, slope of this force-displacement curve, can then be easily extracted, whereas one can also observe the increase of the stiffness with the increase of the asperity width. Stiffness is later plotted along the asperity or contact width (figure 15) and proportionality is confirmed above a critical value 0f 2 mm contact width but there is deviation below that level. The stiffness approaches close to zero with reduced contact width i.e., lowest 0.01 where the nonlinearity is observed.

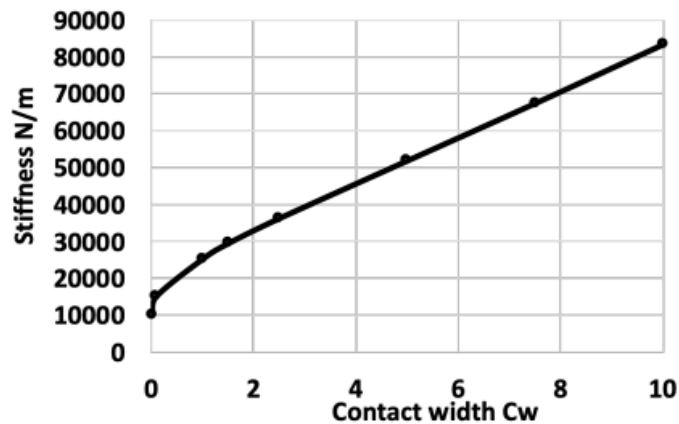

Fig. 15. Resulting contact stiffness versus the asperity/contact width

This is in perfect agreement with the literature from Kendall and Tabor 1971 [36] for value above 2 mm (or close to above 1.5 mm) and gives a valid reason for the assumption that stiffness is proportional to the contact changes. Indeed, it can be concluded that the transmission coefficient is proportional to the contact stiffness in the case of this single asperity where the condition is valid for values above 2 mm (or close to above 1.5 mm) and differs below. This could be due to the fact that although the transmission coefficient is linear to the diameter of the asperity, it cannot be true with stiffness in the lower asperity width regime. To assess the real contact area based on the stiffness, one should know the contact regions and the number of asperities in contact which in reality is difficult. The conclusion from Kendall and Tabor [36] that transmission only depends on the contact stiffness cannot be supported only by a single asperity configuration due to the fact that in the lower stiffness region and lower contact widths, there is a problem of linearity that need to be further studied. Thus, the present work proposes to address this by simulating the wave propagation with multi-asperity contact configurations.

## 9 Establishing the relationship between Wave propagation and interfacial parameters with multiple asperity model

To study the transmission change in the interface, different contact values for given contact width are simulated. In all the changing asperity case the contact width is constant. Thus 25 simulations are performed in covering different interface configurations. As seen in figure 16 from left to right, when the width of the asperity is increased, the resulting number of asperities is reduced too, causing the wave to propagate through asperities of the interface.

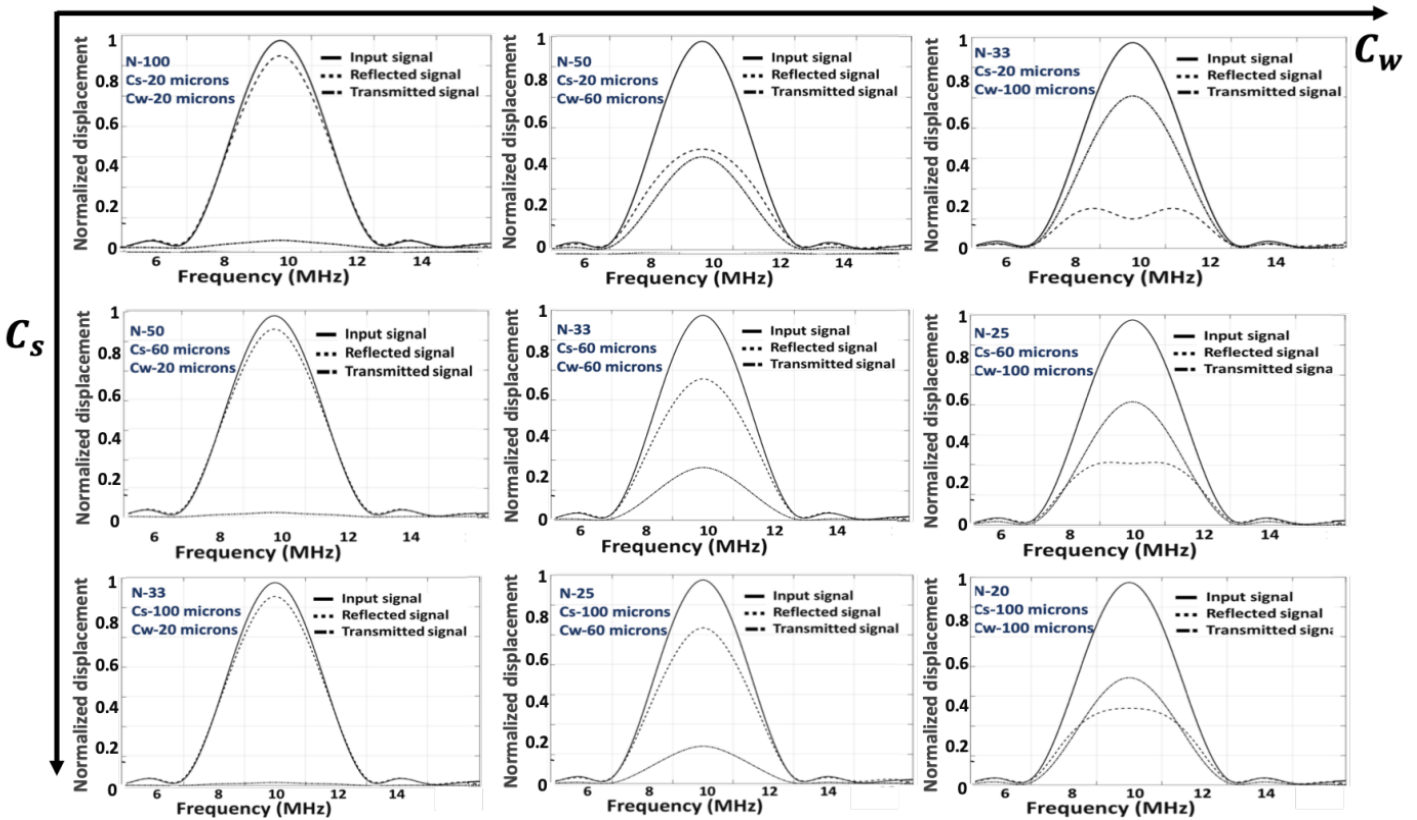

Fig. 16. Compilation of frequency domain plots showing difference in the transmission and reflection coefficients with their respective configuration changes.

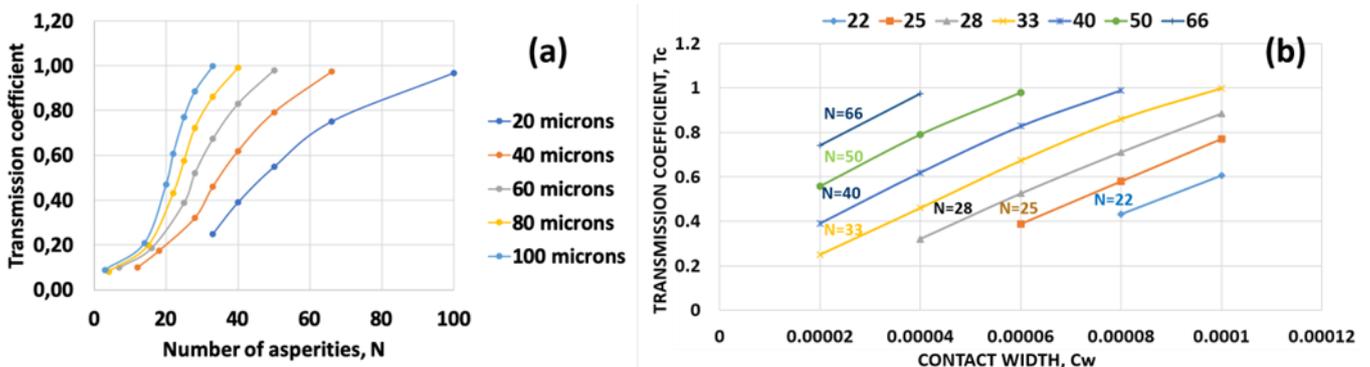

Fig. 17. a) Transmission coefficient plotted in comparison to change in contact width for different number of asperity

configurations, Tc ∝ Cw (b) Transmission coefficient plotted in comparison to contact ratio for different contact width configurationsand number of asperities, Tc /∝ Cr.

As observed from top to bottom, contact space remains consistent for contact width of 20 microns but as the width is increased to 60 microns, reflection increases but with an overall reduced magnitude compared to when contact width is 20 microns. It further reduces to have high transmission in the end. This shows the double dependency of number of asperities and contact width in an interface. To evaluate this dependency, the transmission coefficient is calculated for different interface geometries and comparison plot is drawn for change in number of asperities for all the contact width and contact space configurations. Transmission coefficient is seen monotonically increasing as a function of number of asperities. Figure 17a shows Transmission coefficient plotted along number of asperities; N. linear proportionality can be observed between two parameters until the transmission reaches 0.8 and it is no longer proportional beyond. Individually, Transmission coefficient increases with increases in the asperity number for different contact widths, Cw. This is caused due to change in the interface size and reduction in the air gaps. The results agree with the important hypothesis explained by [9] that, Transmission coefficient, Tc is proportional to number and diameter of the contact. Figure 17b shows the transmission coefficient results obtained by varying the contact width between 20-100 $\mu m$ of the asperity for different values of N. Transmission coefficient increases with increase in the contact width showing a strong linear proportionality. The relationship between the two parameters can be given as, $T_c \propto C_w$. The proportionality is maintained for all the values of N concluding that as the contact width changes, transmission coefficient changes keeping the linear proportionality constant. But each parameter is individually investigated, whereas, the phenomena happen convolved with in the interface. Technically, it's the contact ratio, Cr that acts as the crucial parameter that need to be studied and is defined as,

$$C_r = \frac{Real\ contact\ area}{apparent\ contact\ area} = ENC_w/L \qquad (10)$$

$E$ is young's modulus, L is the length of the apparent contact area, 4 mm in this case and N.Cw are the interface parameters.

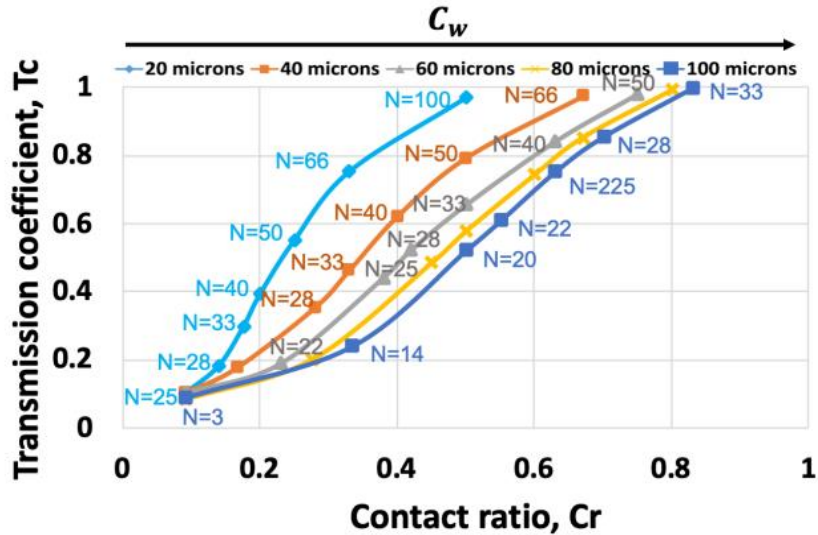

Fig. 18. Transmission coefficient plotted in comparison to contact ratio for different contact width configurations and number of asperities, Tc /∝ Cr.

First thing to be noted in the figure 18 is that transmission coefficient increases with contact ratio, but it is no longer linearly proportional. This is because, for every change in N there is a simultaneous change in Cw. Important thing to be notes is that, for Cr=0.5, there is an increase in the transmission coefficient. Constant contact ratio internally has change in N, Cw. The important aspect that arises in this analysis is, how contact ratio or the interface parameters related to stiffness. Various analytical model was proposed previously to build the relation between interface and wave transmission or reflection. This analysis is in good agreement with the Kendall and Tabor (1971) hypothesis that real contact area is a function of number, size of the asperities in the interface. Most of the research carried out conclude on the theory that stiffness is the sole parameter used in evaluating the real contact area. Although, Dwyer Joyce [17] [12] put focus on the disagreement concluding that real contact area does not just depend on stiffness but also on the other parameters with in the interface

## 10  Relationship between contact stiffness and transmission coefficient

Now that the relationship between the transmission coefficient and other contact parameters is established, it is important to relate the transmission coefficient to the interface stiffness. To do so, a simple stiffness analysis has been performed for each contact configuration previously described in the figure 19. Moreover, complementary contact configurations using the same contact widths but fewer asperities were added in order to extend the range of contact ratios, especially towards the lower values. The contact space has been increased up to 2mm for some configuration.

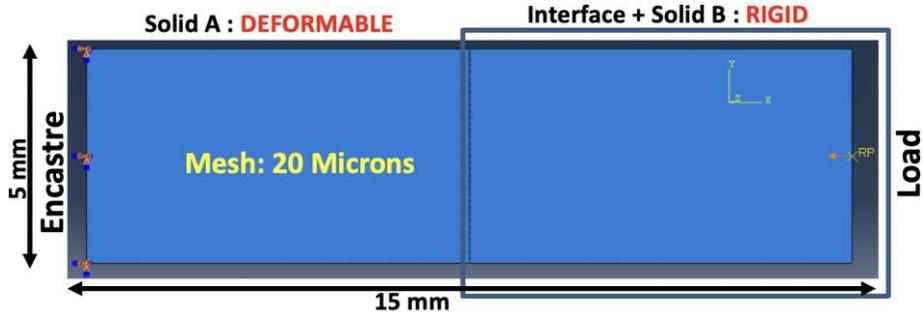

Fig. 19. Geometry corresponding to the model used in determining the interface stiffness.

The selected model is similar in terms of mesh density, material type and interface distribution. However, the study is carried out in static analysis and no wave propagation is implemented. Instead, a small displacement is applied at a reference point and the values of the interface stiffness in each case is extracted based on the reactive force. Various new boundaries and constraints are involved which are shown in figure 19. RP to the right is the reference point at which a 1E-11 amplitude displacement is applied in the negative x direction. A rigid body constraint is applied to both the interface and the solid B making it a combined solid indenting the solid A defined as deformable. Solid A is encastre on the left side ensuring zero displacement. The whole simulation is carried out in ABAQUS/static general case. The simulated reactive force versus displacement curve is extracted and the stiffness value calculated based on the equation:

$$Stiffness, k = (Reactive force)/Input\ displacement\ amplitude \qquad (11)$$

The figure 20 summarizes the evolution of the interface stiffness with the number of asperities for the different contact configurations. It is important to remind that the absolute value of the stiffness has to be carefully considered as it is extracted based on a 2D plane strain model.

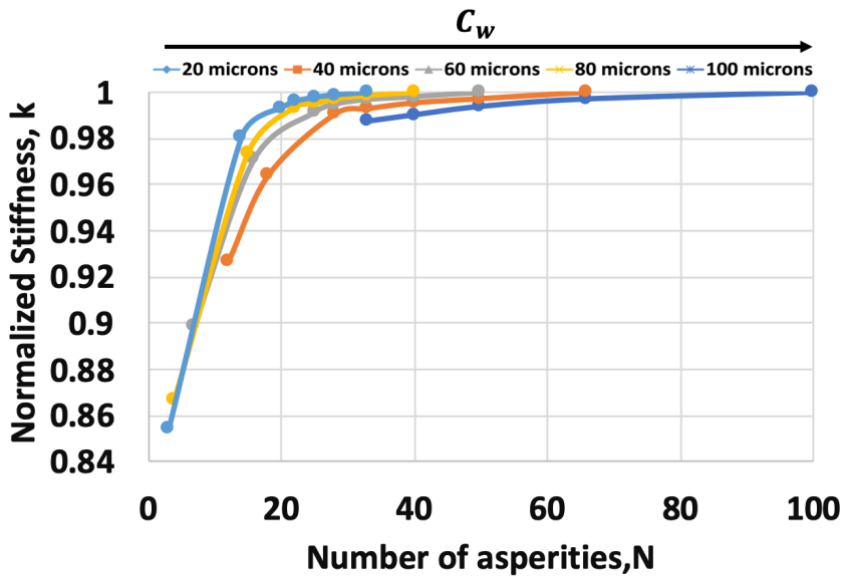

Fig. 20. Relationship between the interface stiffness and the number of asperities.

The observed trends appear to be in agreement with the contact mechanics exposed by [9]: for a given contact width, the stiffness linearly increases with the number of asperities up to a certain point beyond which they start to interact with each other, leading to a slow increase of the stiffness when further increasing N. When focusing on the stiffness computed beyond N = 20, one can still see that a short variation of stiffness from 2 to 4% can occur when increasing the number of asperities or the same order of magnitude when increasing the contact width for a given number of asperities. To conclude, a further relationship is built between stiffness and contact ratio as seen in figure 21.

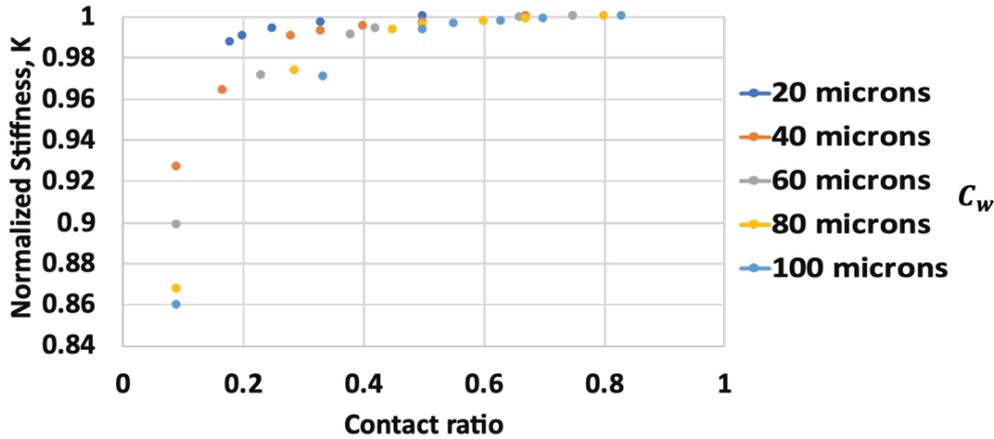

Fig. 21. Relationship between stiffness to contact ratio Included contact ratios in the range of 0.09-0.33.

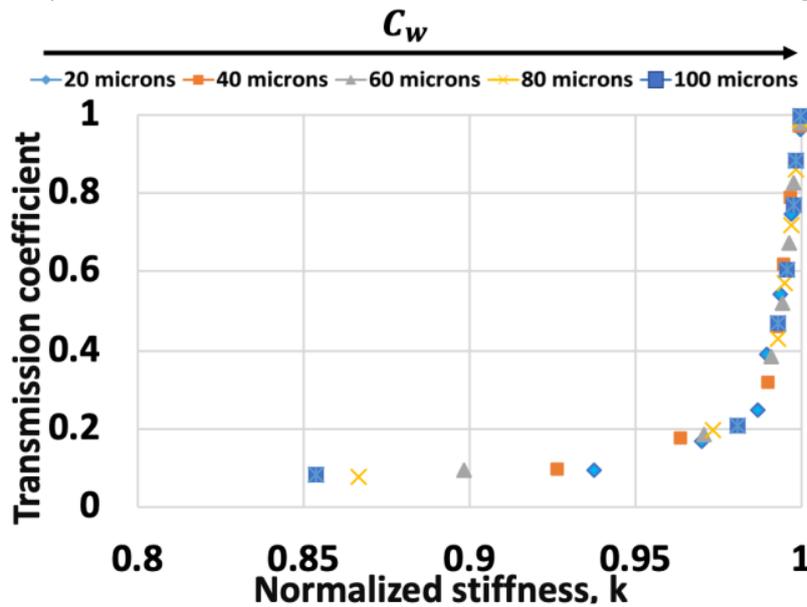

Fig. 22. Transmission coefficient versus the interface stiffness for different contact widths.

When combining this analysis to the wave propagation one, the transmission coefficient can now be plotted versus the interface stiffness (figure 22). It can be noticed that it follows a relatively unique bi-linear evolution with a slope change around the stiffness value of 0.95. Over the lower stiffness value range, i.e., for contact ratios from 0.09 to 0.33, the transmission coefficient is not highly sensitive to a change in stiffness. However, beyond this point, a slight change in stiffness induces a larger variation in terms of transmission coefficient. This could be because at that point, the space between asperities decreases twice to 10 times the value of the previously defined configurations which results in the sudden growth of the transmission curve. It should be important to state that interface stiffness is internally dependent on local interfacial parameters (size, shape and asperity distribution). But also, stiffness could be the same for different Interface configurations. In an interface the asperities even if they initially are far apart, they tend to be very close once new contact points are created, around the existing contacts due to increase in load. This will affect the wave propagation and the influence of local contact parameters. Also, it was previously noted that the transmission coefficient is proportional to the diameter of asperity which was limited to a single asperity. In that analysis, for a single asperity, it was noticed that above 0.5 mm change in diameter is independent it is contact ratio that has to be assessed. But, as the interface deals with multiple asperities, diameter and number are intrinsically dependent, it can in terms of contact ratio be said that above 0.5 different contact ratio combinations will lead to the same stiffness which means the same transmission value. It can be concluded from above analysis that stiffness is partially proportional to transmission coefficient and contact ratio and deviates due to change in internal geometrical parameters that are complex to study. Based on the above-mentioned results, following conclusions are drawn.

## 11  Conclusions

In the paper, a novel numerical approach based on a 2D finite element model was developed to study the interaction between the interface parameters and ultrasonic wave propagation. The transmission coefficient was measured for different model contact configurations and related to the corresponding contact stiffness. This study showed that, when investigating rough contact using a single element transducer, no unique relationship between the interface parameters such as contact width (Cw) or number of asperities (N) and the transmission coefficient could be extracted. Within mind to assess the real contact area in such contacts, it was shown that the contact ratio is also not a unique interface parameter that could be measured using this ultrasonic technique. However, a relationship has been identified when comparing the transmission coefficient to the contact stiffness. If this appears to agree with the literature suggesting a proportionality, it was shown that this seems to depend on the nature of the contact interface, i.e., to depend on the contact ratio. Although being able to measure in-situ the contact stiffness of a given contact is of interest, it was shown that characterizing precisely its morphology, i.e., number of contact spot and their size, required another approach rather to imagine a mapping strategy. Tx part of future study, the transmission coefficient recorded when using a single element transducer could be related to the contact stiffness. However, a sudden change was observed when reaching the largest contact ratio range, i.e., beyond 0.33, where transmission experienced a strong increase. As this phenomenon has not been explained, it would be relevant to develop further analyses to clearly point out the governing mechanisms

## 12  Acknowledgement

This work was supported by the LABEX MANUTECH-SISE (ANR-10 LABX-0075) of Universit de Lyon, within the program" Investissements d'Avenir" (ANR-11-IDEX-0007) operated by the French National Research Agency (ANR).

**References**

[1] S. Wen and P. Huang. *Principles of Tribology*. Wiley, 2012. 1

[2] P.G. Slade. *Electrical Contacts: Principles and Applications, Second Edition*. Electrical and Computer Engineering. Taylor & Francis, 2013. 2

[3] R. Sahli, G. Pallares, C. Ducottet, I. E. Ben Ali, S. Al Akhrass, M. Guibert, and J. Scheibert. Evolution of real contact area under shear and the value of static friction of soft materials. *PNAS*, 115(3):471 – 476, 2018. 2

[4] L. Mourier, D. Mazuyer, A. A. Lubrecht, and C. Donnet. Transient increase of film thickness in micro textured ehl contacts. *Tribology International*, 39:1745 – 1756, 2006. 2

[5] A. de Pannemaecker, J.Y. Buffiere, S. Fouvry, and O. Graton. In situ fretting fatigue crack propagation analysis using synchrotron x-ray radiography. *International Journal of Fatigue*, 97:56 – 69, 2017. 2

[6] J-Y Buffiere, P Cloetens, Wolfgang Ludwig, Eric Maire, and L Salvo. In situ x-ray tomography studies of microstructural evolution combined with 3d modeling. *MRS bulletin*, 33(6):611–619, 2008. 2

[7] Kyle G. Rowe, Alexander I. Bennett, Brandon A. Krick, and W. Gregory Sawyer. In situ thermal measurements of sliding contacts. *Tribology International*, 62:208 – 214, 2013. 2

[8] J. A. Greenwood and J. B. P. Williamson. Contact of nominally flat surfaces. *Proceedings of the Royal Society of London A: Mathematical, Physical and Engineering Sciences*, 295(1442):300–319, 1966. 2

[9] K. Kendall and D. Tabor. An ultrasonic study of the area of contact between stationary and sliding surfaces. *Proceedings of the Royal Society of London. Series A, Mathematical and Physical Sciences*, 323(1554):321–340, 1971. 2, 3, 4, 9, 13, 14

[10] Yoshihisa MINAKUCHI, Kanae YOSHIMINE, Takashi KOIZUMI, and Takanori HAGIWARA. Contact pressure measurements by means of ultrasonic waves : On a method of quantitative measurement. *Bulletin of JSME*, 28(235):40–45, 1985. 2

[11] A. Polijaniuk and J. Kaczmarek. Novel stage for ultrasonic measurement of real contact area between rough and flat parts under quasi-static load. 1993. 3

[12] Jai-Man Baik and R. Bruce Thompson. Ultrasonic scattering from imperfect interfaces: A quasi-static model. *Journal of Nondestructive Evaluation*, 4(3):177–196, 1984. 3, 13

[13] N. F. Haines and D. B. Langston. The reflection of ultrasonic pulses from surfaces. *The Journal of the Acoustical Society of America*, 67(5):1443–1454, 1980. 3

[14] B. Drinkwater, R. Dwyer-Joyce, and P. Cawley. A study of the transmission of ultrasound across real rough solid-solid interfaces. In *1994 Proceedings of IEEE Ultrasonics Symposium*, volume 2, pages 1081–1084 vol.2, Oct 1994. 3

[15] Peter B. Nagy. Ultrasonic classification of imperfect interfaces. *Journal of Nondestructive Evaluation*, 11(3):127–139, 1992. 3

[16] S. Biwa, A. Suzuki, and N. Ohno. Evaluation of interface wave velocity, reflection coefficients and interfacial stiffnesses of contacting surfaces. *Ultrasonics*, 43(6):495 – 502, 2005. 3

[17] R. S. Dwyer-Joyce. The application of ultrasonic NDT techniques in tribology. *Proceedings of the I MECH E Part J Journal of Engineering Tribology*, 219(5):347 – 366, 2005. 3, 13

[18] Bharat Bhushan. Contact mechanics of rough surfaces in tribology: multiple asperity contact. *Tribology Letters*, 4(1):1–35, 1998. 3

[19] J. Krolikowski and J. Szczepek. Prediction of contact parameters using ultrasonic method. *Wear*, 148(1):181 – 195, 1991. 3

[20] Lior Kogut and Izhak Etsion. A finite element based elastic-plastic model for the contact of rough surfaces. *Tribology Transactions*, 46(3):383–390, 2003. 3

[21] K. Komvopoulos and D.-H. Choi. Elastic finite element analysis of multi-asperity contacts. *Journal of Tribology*, 114(4):823–831, October 1992. 3

[22] H G Tattersall. The ultrasonic pulse-echo technique as applied to adhesion testing. *Journal of Physics D: Applied Physics*, 6(7):819, 1973. 3, 4

[23] A.M. Quinn, B.W. Drinkwater, and R.S. Dwyer-Joyce. The measurement of contact pressure in machine elements using ultrasound. *Ultrasonics*, 39(7):495 – 502, 2002. 3

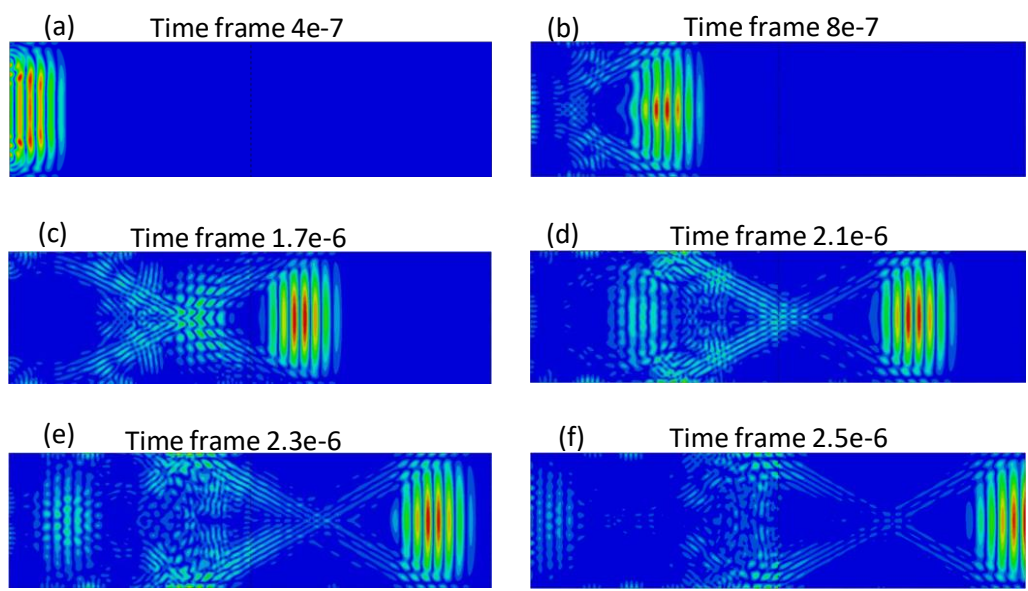

Fig. 23. Wave propagation mode shapes in a solid (10mm) different time intervals showing the propagating and reflected signals.

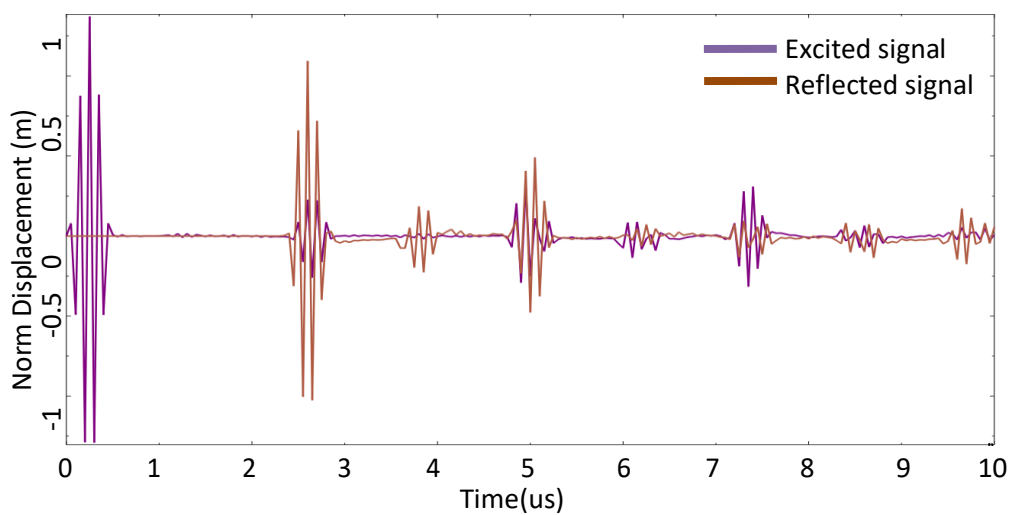

Fig. 24. Excited signal and the reflected signal with multiple reflections shown for the wave propagation through a single solid element.

16